\documentclass[a4paper,11pt]{article}
\pdfoutput=1 

\usepackage{jcappub} 

\usepackage[T1]{fontenc} 
\usepackage[utf8]{inputenc}
\usepackage{babel}

\title{\boldmath  On the virialization threshold for halo mass functions}

\author[a]{Ronaldo C. Batista}
\affiliation[a]{Escola de Ciências e Tecnologia, Universidade Federal do Rio Grande do
Norte, Campus Universitário Lagoa Nova, 59078-970, Natal, RN, Brazil.}

\emailAdd{rbatista@ect.ufrn.br}

\abstract{
In a recent study by Euclid collaboration, the Halo Mass Function (HMF)
has been fitted with accuracy better than $1\%$ for the $\Lambda$CDM
model. Several parameters were introduced and fitted against N-body
simulations, assuming the usual linearly extrapolated matter density contrast
at the collapse time, $\delta_c$, as a basic threshold for halo formation.
As a result, a new function that multiplies $\delta_c$
was introduced, producing an effective threshold that varies both
with redshift and mass scale. We show that the redshift evolution
of this effective threshold is similar to the one of the linear
extrapolated matter density contrast at the virialization time, $\delta_{\rm v}$.
Assuming the Euclid HMF as a fiducial model, we refit the Sheth-Tormen (ST) HMF
using $\delta_{\rm v}$ as a threshold. This new fit improves the agreement between ST-HMF and the Euclid one with respect to Despali et al. (2016) fit, specially at high masses. Interestingly, the parameters
$a$ and $p$ in this refit have values closer to the Press-Schechter
limit of the ST-HMF, showing that the use of $\delta_{\rm v}$
can provide semi-analytical HMF less dependent on extra parameters, while providing better agreement with the Euclid HMF. Moreover, we analyze the
consistency of the ST-HMF fitted with $\delta_{\rm v}$
in smooth dark energy models with time-varying equation of state, finding an overall
good agreement with the evolution of halo abundances expected from the linear evolution of perturbations and the Euclid HMF extrapolated to these scenarios. These findings suggest that
the use of $\delta_{\rm v}$ as a basic function to describe the threshold
for halo formation can be a good guide when considering extrapolations
for models beyond $\Lambda$CDM, which are typically
harder to study in simulations. We also provide
a fitting formula for $\delta_{\rm v}$ in the $\Lambda$CDM model
and a code to compute it for smooth dark energy with equation of state described by
the $w_0w_a$ parametrization.
}

\begin{document}
\maketitle
\flushbottom

\section{Introduction}

Since the first analytical proposal of a Halo Mass Function (HMF)
by Press and Schechter (PS) \cite{Press:1973iz}, several semi-analytical
and numerical studies on the topic have proposed alternative formulations,
e.g., \cite{Sheth1999,Tinker2008b,Bhattacharya2011,Watson2013,Despali2016,OndaroMallea2021,Euclid:2022dbc}.
See \cite{Asgari:2023mej} for an overview of several of these proposals.
One of the key ingredients in the PS formulation, and in many others
based on the Sheth-Tormen (ST) template \cite{Sheth1999}, is the
value of the critical density for the formation of halos, $\delta_{c}$,
which mainly controls the exponential decay of the number of massive
halos. Ideally, this parameter represents the value of the extrapolated
linearly evolved matter contrast, $\delta_{m}$, above which halos
are formed. For many years following the PS proposal, several distinct
values for this threshold were used, e.g., \cite{10.1093/mnras/189.2.203,Colafrancesco1989,Gelb1994}.
Following the suggestions in \cite{Bardeen:1985tr,Bond:1990iw}, the
use of $\delta_{c}\simeq1.686$, the value calculated for Einstein-de-Sitter
(EdS) cosmology using the framework of the Spherical Collapse (SC)
model \cite{Gunn:1972sv,Padmanabhan} at the moment of collapse, became
standard.

In \cite{Sheth:1999su}, a reparametrization of this value was introduced,
such that $\delta_{c}^{2}\rightarrow a\delta_{c}^{2}$, where $a=0.707$.
The value of $a$ can vary when fitted against distinct N-body simulations,
for a compilation of values see \cite{Castro2016}, but $a<1$ is
obtained in almost all cases. This indicates that the usual threshold
for halo formation must be reduced to describe better the halo abundances
seen in numerical simulations. In $\Lambda$CDM cosmology, $\delta_{c}$ is no longer constant. It
is well known that $\delta_{c}$ has a slight decay when $\Lambda$
becomes important for the background evolution \cite{1992ApJ...386L..33L,Lacey1993,Kitayama1996}.
Given this small variation, assuming the constant EdS value of $\delta_{c}$
became usual. 

Recently, the Euclid collaboration has shown that it is possible to obtain and HMF more accurate
than $1\%$ for the $\Lambda$CDM model, at the cost of including
new mass and redshift-dependent functions, \cite{Euclid:2022dbc}
(we refer to this formulation as E-HMF). In particular, the effective
threshold in the E-HMF is given by $a_{R}\Omega_{m}^{a_{z}}(z)\delta_{c}^{2}(z)$,
where $a_{R}$ depends on the mass scale and $\Omega_{m}^{a_{z}}(z)\delta_{c}^{2}(z)$
is redshift-dependent. The value of $a_{z}$ was shown to depend on the halo finder algorithm and ranging between $-0.0693\le a_{z}\le-0.0523$.
Therefore, as $\Lambda$ begins to dominate
the background expansion and $\Omega_{m}(z)$ decays, the effective threshold
increases at low-$z$. Thus, one can understand that the slight decay
of $\delta_{c}(z)$ does not correctly represent the effects of $\Lambda$
on the HMF threshold and the function $\Omega_{m}^{a_{z}}(z)$ compensates
for this decay, introducing an increase in the effective threshold
at low$-z$. This kind of behaviour for the effective threshold was
also found in \cite{Mead2015}, but note that this work aimed to improve
the accuracy of the halo model, not the HMF itself.

As suggested in \cite{Lee_2010,Lee2010b,Lee2010}, one can define another theoretical
threshold for halo formation, $\delta_{{\rm v}}$, associated with virialization time.
The definitions for the collapse and virialization threshold are:
\begin{equation}
\delta_{c}=\delta_{m}^{L}\left(z_{c}\right)\,\,\,\text{and \,\,\,}\delta_{{\rm v}}=\delta_{m}^{L}\left(z_{{\rm v}}\right)\,,
\end{equation}
where $\delta_{m}^{L}$ is the linearly evolved matter density contrast
at the redshift of collapse, $z_{c}$, or virialization, $z_{{\rm v}}$. Both of these quantities
are computed in the framework of the SC model.
For EdS, there is an analytical solution giving a constant $\delta_{{\rm v}}\simeq1.583$.
If one substitutes $\delta_{c}$ by $\delta_{{\rm v}}$ in HMFs, the
threshold is reduced, as effectively done by the values of $a$ in
the ST formulation, indicating that the virialization time can provide
a better description of the abundance of halos, especially at the
high-mass tail of HMFs.

Moreover, the increase of the effective threshold at low-$z$ naturally
occurs for $\delta_{{\rm v}}$ \cite{Batista2017,Chang:2017vhs,Batista:2021uhb}.
Therefore, besides giving a lower baseline value for the effective
threshold, $\delta_{{\rm v}}$ also gives the qualitatively expected
redshift variation, as seen in the E-HMF. Given these two desirable features, it is very
tempting to consider reparametrizations of HMFs using $\delta_{{\rm v}}$
instead of $\delta_{c}$.

This idea was explored in the context of clustering DE \cite{Batista2017},
where the following substitution was proposed $a\delta_{c}^{2}\rightarrow\tilde{a}\delta_{{\rm v}}^{2}$,
where $\tilde{a}=0.803$ is chosen such that the EdS limit of the
ST-HMF is unchanged. The main advantage of this approach is a more
consistent computation of the threshold in the presence of non-matter
fluctuations, see \cite{Batista:2021uhb,Batista:2022ixz} for more
details. Moreover, as explained in \cite{Batista:2022ixz}, is it
not possible to numerically compute $\delta_{c}$ in clustering DE
models with arbitrary sound speed.

In this work, we will further explore the possibility of using $\delta_{{\rm v}}$
instead of $\delta_{c}$ in HMFs, focusing in the $\Lambda$CDM model.
First, we will show that the redshift-dependent part of the E-HMF is in good agreement with
a simple power-law function $\delta_{{\rm v}}(z)$,
with precision better $0.7\%$ for $0.2\le\Omega_{m0}\le0.4$, where $\Omega_{m0} = \Omega_{m}(z=0)$, indicating that the correspondence between the functions is not a
coincidence for a particular value of $\Omega_{m0}$. Then we proceed
with a reparametrization of the ST-HMF with the substitution $\delta_c \rightarrow \delta_{{\rm v}}$ and
assuming the E-HMF as a fiducial model to fit its parameters, showing that this refit agrees
better with E-HMF than the HMF fit given by \cite{Despali2016}. As
a result, we find that the refitted ST parameters $a$ and $p$ get closer
to the values that reproduce the PS-HMF, indicating that the ST-HMF with $\delta_{{\rm v}}$ is less dependent on ad doc parameters. We also analyse how this new version of the ST-HMF behaves when considering smooth DE models with $w\neq-1$, showing it has a good overall agreement with the E-HMF. 
Consequently, one can more confidently use a correction function based on this new version of the ST-HMF to extrapolate the E-HMF to more complex cosmological models, provided that the corresponding modifications are made in the SC or virialization models used to compute $\delta_{\rm v}$.

The plan of this paper is as follows. In section \ref{sec:Thresholds-definitions}
we review and discuss the collapse and virialization definitions for
the thresholds. In section \ref{sec:The-Euclid-fit}, we compare the redshift-dependent
part of the E-HMF with $\delta_{{\rm v}}(z)$. In section \ref{sec:ST-HMF},
substituting $\delta_{c}(z)$ by $\delta_{{\rm v}}(z)$, we refit the ST-HMF,
assuming the E-HMF as a fiducial model. In section \ref{sec:Beyond-CDM-models},
we analyse the consistency of the ST-HMF refitted with $\delta_{{\rm v}}(z)$
for smooth DE models with equation of state described by the $w_{0}w_{a}$ parametrization.
We conclude in section \ref{sec:Conclusions} and provide a representation
of $\delta_{{\rm v}}(z)$ in terms of a well-known fit for $\delta_{c}(z)$
for $\Lambda$CDM model in Appendix A, also indicating a repository
where a code to compute $\delta_{{\rm v}}(z)$ assuming the $w_{0}w_{a}$
parametrization can be found.

Throughout this paper, we assume as baseline the $\Lambda$CDM model
with parameters: $h=0.7$, $\Omega_{c0}h^{2}=0.125$, $\Omega_{b0}h^{2}=0.022$,
$A_{s}=2\times10^{-9}$, $n_{s}=0.965$, $\tau=0.06$, $m_{\nu}=0.06$
and $\Omega_{k0}=0$. In some examples, we vary $\Omega_{m0}$
by changing only $\Omega_{c0}h^{2}$. When analysing DE models
described by the $w_{0}w_{a}$ parametrization, all the other parameters
are kept fixed at the $\Lambda$CDM values. The computation of the
matter power spectrum is made with CAMB \cite{Lewis2000}.

\section{Thresholds definitions\label{sec:Thresholds-definitions}}

Let us first revise how the collapse and virialization thresholds for
the formation of halos are defined and their redshift dependence.
For a more detailed discussion and calculation methods, see \cite{Batista:2021uhb}
and references therein. Both quantities are derived in the framework
of the SC model, which can be summarised by the equations
\begin{equation}
\ddot{R}=-\frac{GM}{R^{2}}\,\,\,\text{and }M=\frac{4\pi}{3}R^{3}\bar{\rho}_{m}\left(1+\delta_{m}\left(t\right)\right)=\text{const},
\end{equation}
where $R$ is the radius of a spherical shell that encloses the conserved
mass $M$, $\bar{\rho}_{m}$ is the background matter density and $\delta_{m}$ the matter density contrast
with no radial variation within the shell.

The usual $\delta_{c}$ threshold is given by the linearly extrapolated
density contrast at the redshift of collapse, $z_{c}$. The collapse
time is associated with the moment at which the radius of the spherical
shell goes to zero, or equivalently when the nonlinearly evolved density
contrast associated with that shell diverges. Thus we have:
\begin{equation}
\lim_{z\rightarrow z_{c}^{+}}\delta_{m}^{N{\rm L}}\left(z\right)=+\infty\,\,\,\text{and }\delta_{c}\left(z_{c}\right)=\delta_{m}^{{\rm L}}\left(z_{c}\right)\,,
\end{equation}
where superscripts $NL$ and $L$ represent the nonlinear and linear
evolutions of $\delta_{m}$, respectively.

The virialization threshold, introduced in \cite{Lee_2010,Lee2010b,Lee2010},
is given by the linearly extrapolated density contrast at the redshift
of virialization, $z_{{\rm v}}$,
\begin{equation}
\delta_{{\rm v}}\left(z_{{\rm v}}\right)=\delta_{m}^{L}\left(z_{{\rm v}}\right).
\end{equation}
The moment of virialization is also given by the SC model, when virial equilibrium is achieved. Following \cite{Basse2012}, the
virialization time is determined when the equation below is satisfied
\begin{equation}
\frac{1}{R^2}\left(\frac{dR}{dt}\right)^2 + \frac{1}{R}\frac{d^2 R}{dt^2}=0
\end{equation}
In EdS, the moment of virialization is such that we have 
\begin{equation}
\delta_{{\rm v}}\simeq1.583\,\,\,\text{and }\Delta_{{\rm v}}=\frac{\rho_{m}\left(z_{{\rm v}}\right)}{\bar{\rho}_{m}\left(z_{{\rm v}}\right)}\simeq146.8,\,
\label{deltas-vir}
\end{equation}
where $\rho_{m}\left(z\right)=\bar{\rho}_{m}\left(z\right)\left[1+\delta_{m}^{NL}\left(z\right)\right]$
is the total matter density. Note that $\Delta_{{\rm v}}$ is not
the usual virialization overdensity, which is given by $\Delta=\rho_{m}\left(z_{\rm v}\right)/\bar{\rho}_{m}\left(z_{c}\right)\simeq177.7\,.$ It is also worth mentioning that the spherical overdensity used to identify halos in the E-HMF analysis has a similar behaviour to the virialization overdensity defined in (\ref{deltas-vir}), see quantities $\Delta_{cc}$ and $\Delta_{v}$ in figure 1 of reference \cite{Batista:2021uhb}.

It is interesting to note that the virialization equation can be generalised
to include other components, such as smooth or clustering DE \cite{Mota:2004pa,Maor:2005hq,Basse2012,Batista2017,Chang:2017vhs},
whereas the collapse definition is mostly insensitive to other ingredients
that participate in the collapse process. Though in general $\delta_{m}^{NL}$
depends on the background evolution and other clustering components,
near the collapse time, it is growing nearly exponentially, responding
essentially to its own nonlinear value. Thus, the background evolution,
and possibly other subdominant fluctuations, have very small impact
on the value of $\delta_{c}(z)$, typically less than $1\%$ with respect
to the EdS value \cite{Batista:2021uhb}.

\begin{figure}
\centering{}
\includegraphics[scale=0.6]{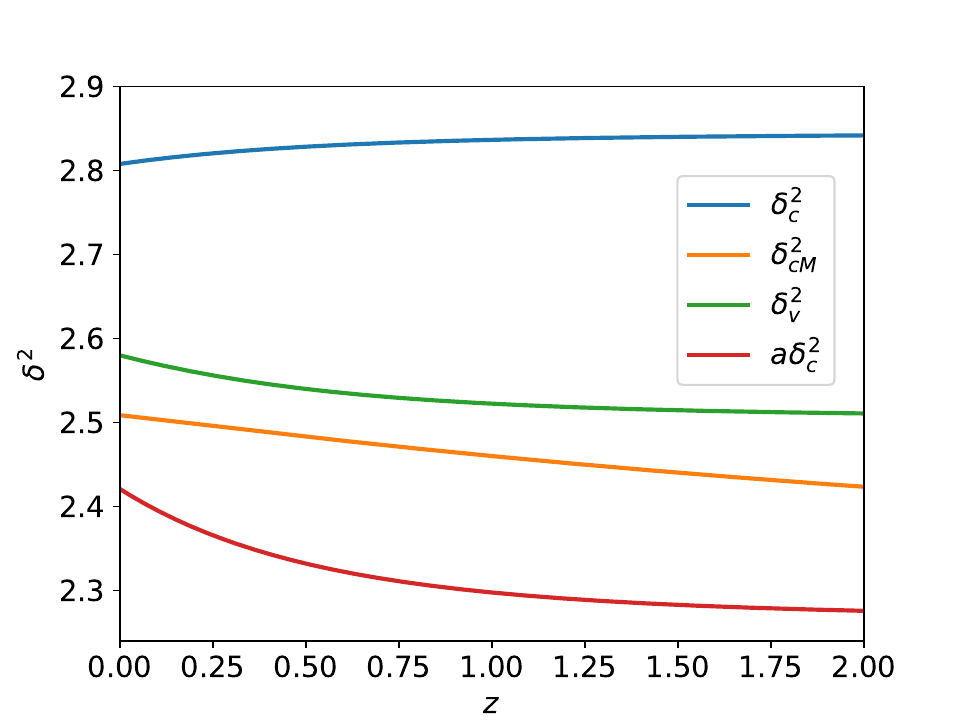}
\caption{Evolution of $\delta_{c}^{2}(z)$, $\delta_{cM}^{2}(z)$, $\delta_{{\rm v}}^{2}(z)$
and $a(M,z)\delta_{c}^{2}(z)$ as a function of redshift in the $\Lambda$CDM
model with $\Omega_{m0}=0.3$, $a_{z}=-0.0658$ and $a_{R}=0.7965$
(corresponding to a mass-scale $10^{14}M_{\odot}h^{-1}$ ). 
\label{fig:deltas}}
\end{figure}

Summarising, in the EdS model, the two thresholds can be analytically
computed and are constants: 
\begin{equation}
\delta_{c\,\text{EdS}}\simeq1.686\,\,\,\text{and}\,\,\,\delta_{{\rm v\,\text{EdS}}}\simeq1.583\,.
\end{equation}
When DE begins to dominate the expansion, they become redshift-dependent.
Whereas $\delta_{c}(z)$ has a slight decay when DE becomes important,
$\delta_{{\rm v}}(z)$ presents a mild growth, see figure \ref{fig:deltas}.

In this context, it is also interesting to mention that the following
threshold
\begin{equation}
\delta_{cM}(z)=1.59+0.3114\times\ln\sigma_{8}\left(z\right)\,,\label{eq:delta_c-Mead}
\end{equation}
proposed in \cite{Mead2015} to improve the accuracy of the nonlinear
power-spectrum computed via the halo model, also has smaller values
than $\delta_{c}(z)$ and a mild growth at low$-z$. In figure \ref{fig:deltas}
we show the evolution of the squares of $\delta_{c}(z)$, $\delta_{{\rm v}}(z)$,
$\delta_{cM}(z)$ and the effective
threshold in the E-HMF, $a(M,z)\delta_{c}^{2}(z)$, which will discuss in
more detail in section \ref{sec:The-Euclid-fit}. In this work, we compute  $\delta_{c}(z)$ using the fit given in \cite{Kitayama1996} and $\delta_{{\rm v}}(z)$ numerically, with the code available at \href{https://github.com/roncab/scollas}{https://github.com/roncab/scollas}.  

Assuming that the best representation of the effective threshold is given by $a(M,z)\delta_{c}^{2}(z)$,
where $a(M,z)$ is given by (\ref{eq:a-parametrization}), we clearly see
that the usual $\delta_{c}(z)$ is both higher and has the opposite redshift
variation. On the other hand, $\delta^{2}_{{\rm v}}(z)$ is closer to
$a(M,z)\delta^{2}_{c}(z)$ and also increases at low-$z$. Note that the fit given in (\ref{eq:delta_c-Mead}) was obtained in order to improve the accuracy of the halo model and the corresponding nonlinear matter 
power spectrum computed with it, and not necessarily the HMF, which is the quantity of interest in this work. But it is also clear that its behaviour is more compatible with $a(M,z)\delta_{c}^{2}(z)$ than $\delta_{c}^{2}(z)$. 

In principle, $\delta_{cM}(z)$ also has good agreement with the effective threshold in the E-HMF. However, as can be seen in figure \ref{fig:deltas}, the time variation of $\delta_{\rm v}(z)$ is more compatible with $a(M,z)\delta_{c}^{2}(z)$ than $\delta_{cM}(z)$. In particular, the $\ln\sigma_{8}\left(z\right)$ dependence in $\delta_{cM}(z)$ will not give a constant value at high-z, as should be in the EdS limit. Moreover, $\delta_{\rm v}(z)$ is described by a physical model, which can be improved and generalised for other cosmologies,  whereas $\delta_{cM}(z)$ is based on a empirical fit. Note also that, in figure \ref{fig:deltas}, the bare value of $\delta_{\rm v}(z)$ is shown, but $\delta_{cM}(z)$ has two adjusted parameters, so it is closer to the E-HMF effective threshold.

\section{The effective threshold in the E-HMF and $\delta_{{\rm v}}$ \label{sec:The-Euclid-fit}}

In this section, we analyse how the effective threshold of E-HMF is
related to $\delta_{{\rm v}}(z)$ and define the main quantities of interest.
The comoving number density of halos per mass interval is given by
\begin{equation}
\frac{dn}{dM}=\frac{\rho_{m}}{M}\nu f\left(\nu\right)\frac{d\ln\nu}{dM}\,,
\end{equation}
where $\nu f\left(\nu\right)$ is the multiplicity function, $\nu=\delta_{c}(z)/\sigma$,
$\sigma ^2$ is the smoothed variance of the matter power-spectrum, $P\left(k\right)$,
given by 
\begin{equation}
\sigma^{2}\left(R,z\right)=\frac{1}{2\pi^{2}}\int dkk^{2}W^{2}\left(kR\right)P\left(k,z\right)\,,
\end{equation}
$W\left(kR\right)$ is the Fourier transform of the top-hat window
function and $R=\left(\frac{3M}{4\pi\bar{\rho}_{m}}\right)^{1/3}$.

The base form of the HMF fitted by the Euclid collaboration was proposed
in \cite{Bhattacharya2011} and is given by

\begin{equation}
\nu f\left(\nu\right)_{E}=A\left(p,q\right)\sqrt{\frac{2a\nu^{2}}{\pi}}\left(1+\frac{1}{\left(a\nu^{2}\right)^{p}}\right)\left(\sqrt{a}\nu\right)^{q-1}\exp\left(-\frac{a\nu^{2}}{2}\right)\,.\label{eq:HMF-euclid}
\end{equation}
The Euclid analysis has assumed that the parameters $A,\, p,\, q$ and $a$
have the following forms:
\begin{equation}
A\left(p,q\right)=\left\{ \frac{2^{-1/2-p+q/2}}{\sqrt{\pi}}\left[2^{p}\Gamma\left(\frac{q}{2}\right)+\Gamma\left(-p+\frac{q}{2}\right)\right]\right\} ^{-1}\,,\label{eq:Bha-normalization}
\end{equation}

\begin{equation}
a(M,z)=a_{R}\Omega_{m}^{a_{z}}(z)=\left[a_{1}+a_{2}\left(\frac{d\ln\sigma}{d\ln R}+0.6125\right)^{2}\right]\Omega_{m}^{a_{z}}(z)\,,\label{eq:a-parametrization}
\end{equation}
\begin{equation}
p(M,z)=p_{1}+p_{2}\left(\frac{d\ln\sigma}{d\ln R}+0.5\right)\,,
\end{equation}
\begin{equation}
q(M,z)=\left[q_{1}+q_{2}\left(\frac{d\ln\sigma}{d\ln R}+0.5\right)\right]\Omega_{m}^{q_{z}}(z)\,,
\end{equation}
where $a_{1}$, $a_{2}$, $a_{z}$, $p_{1}$, $p_{2}$, $q_{1}$,
$q_{2}$ and $q_{z}$ are constants fitted with several N-body simulations
discussed in \cite{Euclid:2022dbc}. The function $A\left(p,q\right)$
normalises the HMF when $p$ and $q$ are constants. 

As can be seen, the function $a(M,z)$ always multiplies $\delta_{c}^{2}(z)$
and depends both on redshift and mass scale. The usual SC model is scale-independent,
therefore it is not able to explain the variation with mass scale,
which is encoded in $a_{R}$. Thus, we will focus on the redshift
dependence of the effective threshold, $\Omega_{m}^{a_{z}}(z)\delta_{c}^{2}(z)$,
and analyse its relation with $\delta_{{\rm v}}(z)$.

The evolution of $a(M,z)\delta_{c}^{2}(z)$ is shown in figure \ref{fig:deltas},
where we used $a_{R}=0.7965$, corresponding to a mass of $10^{14}M_{\odot}h^{-1}$.
We can see that both $\delta_{{\rm v}}^{2}(z)$ and $a(M,z)\delta_{c}^{2}(z)$
increase at low-$z$, so they qualitatively represent the same trend
for halo formation, i.e., a threshold that increases as $\Lambda$
becomes more important. Thus it is clear that the term $\Omega_{m}^{a_{z}}(z)$
inside $a(M,z)$ compensates the slight decay of $\delta_{c}(z)$ to better
describe the abundance of simulated halos. Moreover, $\delta_{{\rm v}}^{2}(z)$
is closer to $a(M,z)\delta_{c}^{2}(z)$ then $\delta_{c}^{2}(z)$. Therefore,
also from a direct analysis of the E-HMF, one can conclude that the
physics associated with the definition of $\delta_{{\rm v}}(z)$ is more
suitable than $\delta_{c}(z)$ because it is closer to the
effective threshold and grows at low-$z$. 

\subsection*{Relation between the effective threshold and $\delta_{\rm v}$}

Although $\delta_{\rm v}^2(z)$ and  $a(M,z)\delta_{c}^{2}(z)$  have similar behaviours, it is also clear that there exists a mismatch between them. In figure \ref{fig:deltas}, we can see that $a(M,z)\delta_{c}^{2}(z)$ grows faster at late times and has a lower value in the EdS limit. Therefore, it is important to check whether the similarity between them is not a coincidence for a specific set of cosmological parameters. Our goal now is
to find a representation of the redshift-dependent part of the E-HMF effective threshold, given by $\Omega_{m}^{a_{z}}(z)\delta_{c}^{2}(z)$, in
terms of $\delta_{{\rm v}}(z)$ and later check whether it is robust
against variations of $\Omega_{m0}$.  Let us analyse the following
parametrization
\begin{equation}
\Omega_{m}^{a_{z}}(z)\delta_{c}^{2}(z)=\alpha\delta_{{\rm v}}^{\beta}\left(z\right)\,,\label{eq:remap-1}
\end{equation}
where $\alpha$ and $\beta$ are constants. 

It is important to note
that the form $\alpha\delta_{{\rm v}}^{\beta}\left(z\right)$ will introduce a non-standard power-law dependence on the density threshold, i.e., distinct from power $2$. This is unusual in the context of statistical derivation of HMF \cite{Press:1973iz,Bardeen:1985tr},
because the squared filtered density contrast has different power
than its variance. However, this reparametrization will not be implemented in the HMFs analysis in the next sections and will be used only to check the consistency between the functions when varying $\Omega_{m0}$. 

Two other forms preserving the power two in $\delta_{\rm v}$ that could be used are: $\alpha\delta_{\rm v}^{2}(z)$ and $\alpha\Omega_{m}^{\beta}(z)\delta_{{\rm v}}^{2}\left(z\right)$. We checked that both of them are very precise when fitted for a fixed $\Omega_{m0}=0.3$, but the resulting fits are not robust when extrapolated to other values of $\Omega_{m0}$. Therefore, we will proceed with the expression in equation (\ref{eq:remap-1}), which will bring an important insight about the virialization model.

The constant $\alpha$ can be set by assuming that $\Omega_{m}^{a_{z}}(z)\delta_{c}^{2}(z)=\alpha\delta_{{\rm v}}^{\beta}(z)$
in the EdS limit, thus
\begin{equation}
\alpha=\frac{\delta_{c\,\text{EdS}}^{2}}{\left(\delta_{{\rm v}\,\text{EdS}}\right)^{\beta}}\,,\label{eq:alpha1-definition}
\end{equation}
and $\beta$ is given by
\begin{equation}
\beta=\ln\left(\frac{\Omega_{m}^{a_{z}}(z)\delta_{c}^{2}\left(z\right)}{\delta_{c\,\text{EdS}}^{2}}\right)/\ln\left(\frac{\delta_{{\rm v}}\left(z\right)}{\delta_{{\rm v}\text{\,EdS}}}\right)\,,
\end{equation}
which varies with $z$. For $\Omega_{m0}=0.3$ and $0\le z\le2$,
we have $4.05<\beta<4.51$.

One could not expect that these two representations would match exactly.
Nevertheless, we can compute an effective $\beta$ that minimises
the difference between them for some redshift interval of interest.
We choose to minimise 
\begin{equation}
\chi^{2}\left(\beta\right)=\sum_{z}\left[\Omega_{m}^{a_{z}}(z)\delta_{c}^{2}\left(z\right)-\delta_{c\,\text{EdS}}^{2}\left(\frac{\delta_{{\rm v}}\left(z\right)}{\delta_{{\rm v}\,\text{EdS}}}\right)^{\beta}\right]^{2}\,.\label{eq:xi2}
\end{equation}
taking the sum in redshift from $z=0$ to $z=2$ with $10$ steps.
We verified that concentrating the $z$ interval at lower $z$ does
not change the results significantly. For $\Omega_{m0}=0.3$, we
get
\begin{equation}
\alpha=0.3816\text{ and }\beta=4.3745\,.
\label{eq:alph-beta-fit}
\end{equation}

\begin{figure}
\centering{}\includegraphics[scale=0.6]{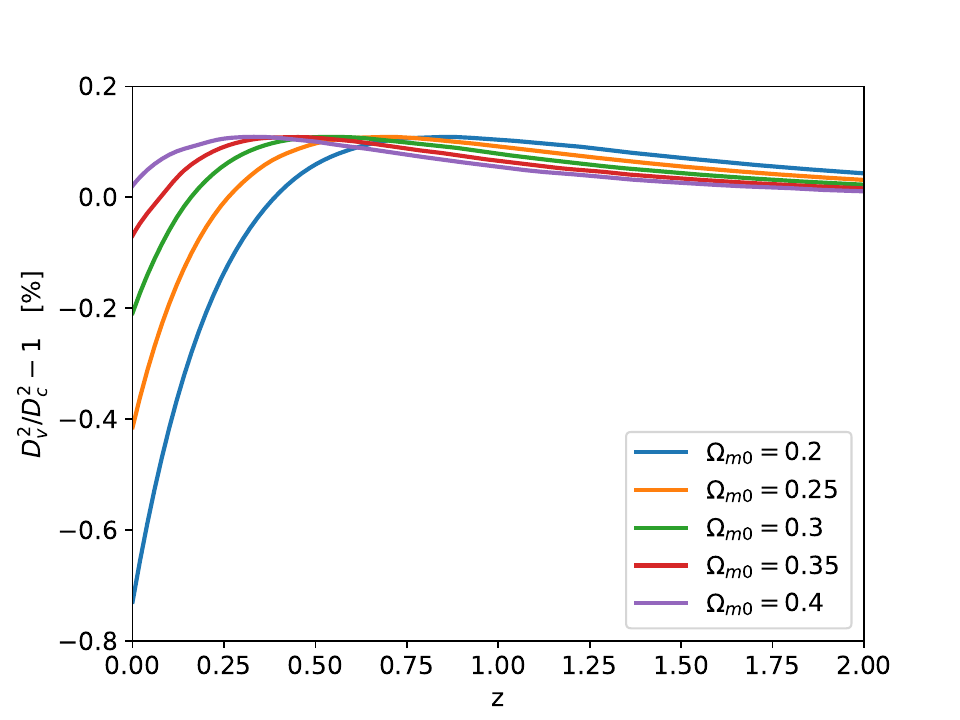}
\caption{Difference between $D_{{\rm v}}^{2}$ and $D_{c}^{2}$ defined in
(\ref{eq:definition-Dc-Dv}) for several values of $\Omega_{m0}$
indicated in the legend for $\Lambda$CDM model.
\label{fig:Dv-error}}
\end{figure}

Now let us see the level of agreement between the two representations.
For convenience, we define:
\begin{equation}
D_{c}^{2}\left(z\right)\equiv\Omega_{m}^{a_{z}}(z)\delta_{c}^{2}\left(z\right)\,\,\,\text{and}\,\,\,D_{{\rm v}}^{2}\left(z\right)\equiv\alpha\text{\ensuremath{\delta_{{\rm v}}^{\beta}\left(z\right)}}\,,
\label{eq:definition-Dc-Dv}
\end{equation}
where $\alpha$ and $\beta$ are given by (\ref{eq:alph-beta-fit}).
In figure \ref{fig:Dv-error}, we show the percent differences between
$D_{{\rm v}}^{2}$ and $D_{c}^{2}$ for several values of $\Omega_{m0}$.
We observe that the difference increases for lower $\Omega_{m0}$
values. This is expected because, when determining the parameter $\alpha$,
we assumed EdS-limit, which is not well satisfied for low $\Omega_{m0}$.
On the other hand, the difference decreases for higher values of $\Omega_{m0}$.

As can be seen, the redshift-dependent part
of the effective threshold in the E-HMF is well described by a power-law
function of $\delta_{{\rm v}}$, even for values of $\Omega_{m0}$
distinct from the value used to determine the values of $\alpha$
and $\beta$. The fact that $\Omega_{m}^{a_{z}}(z)\delta_{c}^{2}\left(z\right)\propto\delta_{{\rm v}}^{4.37}(z)$
can be understood as manifestation of the non-universality of the E-HMF, introduced
by the new fitting functions. Another possible interpretation is that a more realistic SC model should provide a virialization threshold that grows faster at low-$z$, in such a way that the new function could represent $D^2_c(z)$ with $\beta \simeq 2$, but this analysis is beyond
the scope of this paper.

In order to get a HMF less dependent on empirical fitting function encoded in $D_c$, one could try to use the substitution
$D_c \rightarrow D_{\rm v}$ in the E-HMF. As figure \ref{fig:Dv-error} indicates, this kind of reparametrization can be very accurate. We checked that it produces less than $5\%$ differences against the original E-HMF for $10^{13} M_{\odot}/h \le M \le 5\cdot10^{15} M_{\odot}/h$, $0.25 \le \Omega_{m0} \le 0.35$ and $0\le z \le2$. However, we would be dealing with a reparametrization of a non-universal HMF, whose applicability to other cosmological models would be quite speculative. Therefore, in what follows, we will study how the use of $\delta _{\rm v}$ in the more universal ST-HMF can provide a better agreement with the E-HMF and later analyse how it can used to extrapolate the E-HMF to models beyond $\Lambda$CDM.

\section{Sheth-Tormen halo mass function with $\delta_{{\rm v}}$\label{sec:ST-HMF}}

Given that the redshift-dependent part of the effective threshold
in the E-HMF is qualitatively similar to $\delta_{{\rm v}}$,
we might ask whether the use of the virialization threshold provides
a better agreement between semi-analytic HMFs and the E-HMF, which
currently is the most accurate description of halo abundance for
the $\Lambda$CDM model. Let us consider the ST-HMF \cite{Sheth1999},
which has the form
\begin{equation}
\nu f\left(\nu\right)_{ST}=A\sqrt{\frac{2a\nu^{2}}{\pi}}\left(1+\frac{1}{\left(a\nu^{2}\right)^{p}}\right)\exp\left(-\frac{a\nu^{2}}{2}\right)\,,\label{eq:ST-HMF}
\end{equation}
where $a$ and $p$ are constant fitted parameters, $A\left(p\right)=\left[1+2^{-p}\Gamma\left(1/2-p\right)/\sqrt{\pi}\right]^{-1}$
when assuming that the ST-HMF is normalised, or can be fitted as well.

The original values of the parameters are $a=0.707$, $p=0.3$ and
$A=0.322$. Besides many other refits of this HMF, more recently \cite{Despali2016}
produced a fit using the virialization overdensity given by \cite{Bryan:1997dn},
showing that identifying halos with this overdensity function provides
a more universal HMF. The result for a fit considering all redshifts
and models considered in that work is: 
\begin{equation}
a=0.7989\,,\,\,\,p=0.2536\,,\,\,\,A=0.3295\,.
\label{eq:despali-par}
\end{equation}
The comparison made in \cite{Euclid:2022dbc} between their HMF and
this fit indicates that, at $z=0$, the two functions disagree by
less than $5\%$ up to roughly $4.8\times10^{14}M_{\odot}h^{-1}$
and by less than $10\%$ up to $7\times10^{14}M_{\odot}h^{-1}$. The
differences increase at $z=1$ and $z=2$.

Now let us refit these parameters substituting $\delta_{c}(z)$
by $\delta_{{\rm v}}(z)$. First, we mention that a simple way to substitute
$\delta_{c}$ by $\delta_{{\rm v}}$ was given \cite{Batista2017},
where the original parameter $a=0.707$ was rescaled such that the
EdS limit is the same, so
\begin{equation}
a\rightarrow\tilde{a}=a\frac{\delta_{c\,\text{EdS}}^{2}}{\delta_{{\rm v}\,\text{EdS}}^{2}}\simeq0.803\,,\label{eq:a-refit-17}
\end{equation}
maintaining $p=0.3$ and $A\left(p\right)=\left[1+2^{-p}\Gamma\left(1/2-p\right)/\sqrt{\pi}\right]^{-1}$
as a normalisation factor. We have verified that the reparametrization
given by (\ref{eq:a-refit-17}) deviates much more from the E-HMF
than the one given by \cite{Despali2016}, equation (\ref{eq:despali-par}).
This deviation is mainly due to the high value of $p$ used. As we
will see, the preferred values of $p$ with the virialization threshold
is close to zero.

Assuming the E-HMF as fiducial HMF representation, we determine the new
set of parameters by minimising 
\begin{equation}
\chi^{2}\left(a,p,A\right)=\sum_{M,z}\left(\nu f\left(\nu\right)_{|ST{\rm v}}-\nu f\left(\nu\right)_{|E}\right)^{2}\,,
\label{eq:xi2-ST-HMF}
\end{equation}
where $\nu f\left(\nu\right)_{|E}$ is given by (\ref{eq:HMF-euclid})
and STv denotes (\ref{eq:ST-HMF}) with the substitution $\delta_{c}\rightarrow\delta_{{\rm v}}$.
We take the mass summation between $10^{14}M_{\odot}h^{-1}$ and $10^{15}M_{\odot}h^{-1}$with
10 log-spaced steps and the redshift summation with $0\le z\le2$
in 10 steps. Note that this simple definition gives more weight to
the differences at the peak of HMF at each redshift. The minimisation
of (\ref{eq:xi2-ST-HMF}) indicates 
\begin{equation}
a=0.8978\,,\,\,\,p=0.0031\,,\,\,\,A=0.3184\,.\label{eq:ST-delta-vir-fit}
\end{equation}

We have verified that modifying the mass range in (\ref{eq:xi2-ST-HMF})
to $10^{13}-10^{15}M_{\odot}h^{-1}$ with 10 log-spaced steps has
small impact on the best-fit parameters, with a slight worsening of
the fit. The range used for the determination of (\ref{eq:ST-delta-vir-fit})
was chosen for better accuracy at the high-mass tail, which is very
sensitive to the cosmological model.

It is interesting to note that the PS limit of the ST-HMF is given
by
\begin{equation}
a\rightarrow1\,,\,\,\,p\rightarrow0\,,\,\,\,A\rightarrow1/2\,.
\end{equation}
Therefore, as can be seen from the values in (\ref{eq:ST-delta-vir-fit}),
refitting the ST-HMF with $\delta_{{\rm v}}$ produces parameters $a$
and $p$ that are much closer to the PS values than the original values
or the more recent fit (\ref{eq:despali-par}), while the parameter
$A$ has a very small change. Moreover, even when fitting for larger
mass intervals, the parameters $a$ and $p$ still reach values closer
to the PS limit when compared to the values of (\ref{eq:despali-par}). 
Since the PS-HMF is an analytical formulation,
this a concrete example that the use of $\delta_{{\rm v}}$ as basic threshold function can provide
a HMF less dependent on ad doc parameters, while providing
better agreement with the E-HMF.

\begin{figure}
\centering{}\includegraphics[scale=0.5]{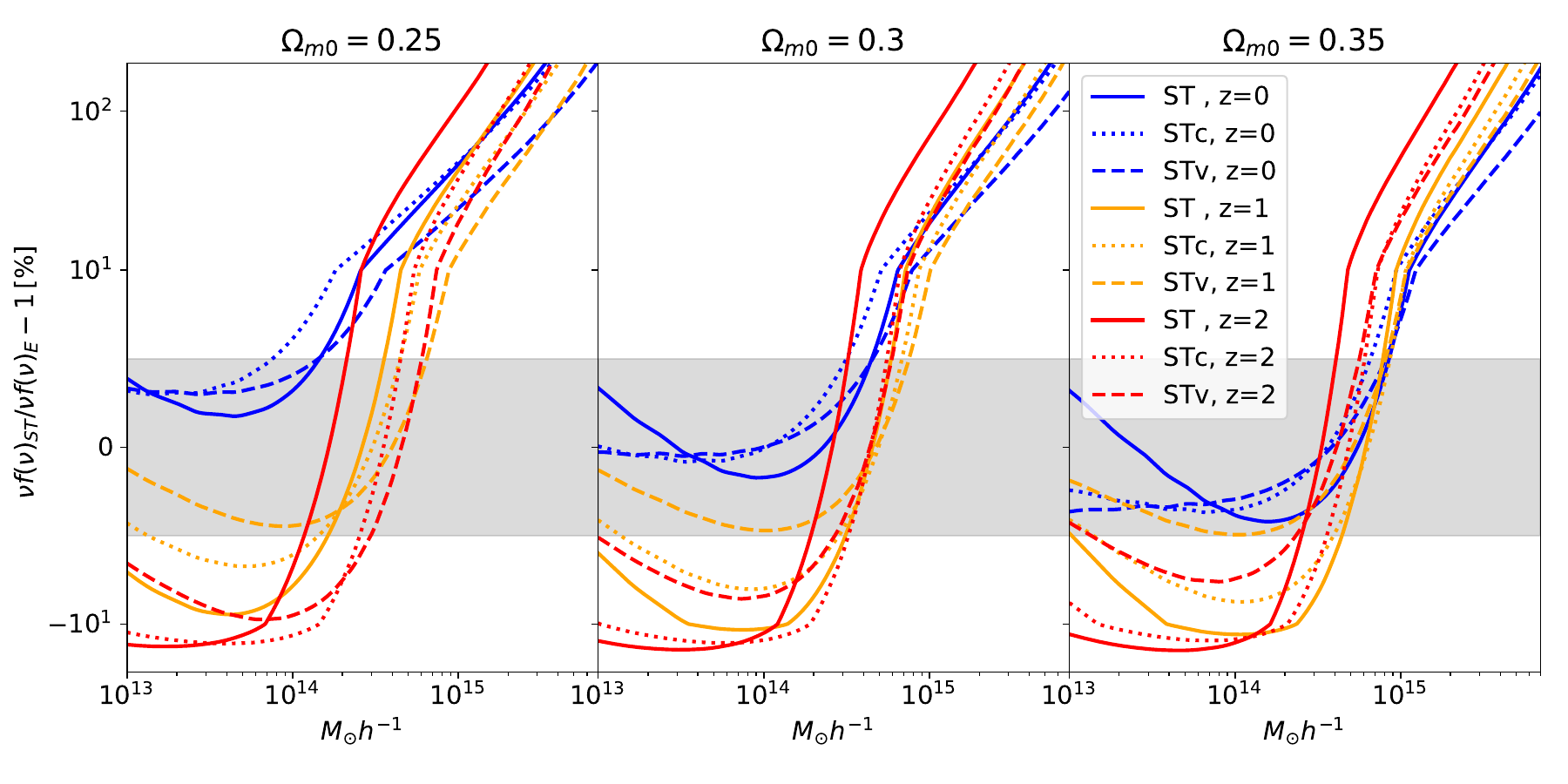}
\caption{Differences between ST template with respect to the E-HMF, equation (\ref{eq:HMF-euclid}), for three sets o parameters. The ST label indicates the parameters given by Despali et al. 2016, ref. \cite{Despali2016}, (solid lines), equation (\ref{eq:despali-par}). The STv label indicates our refit using  
$\delta_{{\rm v}}$ (dashed lines), equation (\ref{eq:ST-delta-vir-fit}), and STc the refit of the ST template using $\delta_{c}$, done in the same fashion as for the STv case (dotted lines). Each panel shows the
differences for distinct values of $\Omega_{m0}$, indicated in
the title of each plot, and for three redshifts indicated in the legend
of the right panel. The y-scale is linear between $\pm10\%$ and symmetric-log beyond these values. The gray bands indicate $\pm5\%$ region. 
\label{fig:ST-error}}
\end{figure}

In figure \ref{fig:ST-error}, we show the percent differences of ST and STv-HMFs
against the E-HMF. The general trend is clear:
the STv-HMF has better agreement with E-HFM than ST-HMF,
expect for some small intermediate mass interval. In particular, at
$z=0$ and for $\Omega_{m0}=0.3$, the deviation of STv from E-HMF
is smaller than $1\%$ for $M<1.8\times10^{14}M_{\odot}h^{-1}$ and
for $M>4.4\times10^{14}M_{\odot}h^{-1}$ it is always smaller then
ST-HMF.

Besides the improved agreement with the E-HMF, it is also important
to note that, as expressed in \cite{Despali2016}, the residuals between 
their model and  their numerical HMF ``are larger at high $\nu$, where the
best fit predicts more halos than we find in the simulations at $z=0$.''
Since $\delta_{{\rm v}}(z)$ increases at low-$z$, whereas $\delta_{c}(z)$
decreases, it is clear that a HMF with the virialization threshold
is more suitable. This is also represented by the better agreement
between STv and E-HMFs at higher masses, which occurs for all redshifts
and values of $\Omega_{m0}$ shown in figure \ref{fig:ST-error}.

One might ask\footnote{We thank the referee for point this issue.} whether the agreement between ST template (still assuming $\delta_c(z)$ as a threshold) and E-HMF can be improved if it is refitted in the same fashion as we did for the STv-HMF, i.e., assuming the E-HMF as a fiducial model and minimising the analogous $\chi^2$, given by equation (\ref{eq:xi2-ST-HMF}).
This refit indicates:
\begin{equation}
a=0.7854\,,\,\,\,p=0.0990\,,\,\,\,A=0.3195\,.
\label{eq:ST-crit-new}
\end{equation}
Note that the main difference with respect to the Despali et al. fit, equation (\ref{eq:despali-par}), is lower $p$ value. The main effect of this new set of parameters is a better agreement between ST and E-HMFs at low masses. In figure \ref{fig:ST-error}, we refer to this new fit as STc. As can be seen, the behaviour at high masses is similar to the one for ST-HMF. When compared with this refit, the STv-HMF still performs better, especially at high masses. In particular, we can compare the values of $\chi^2$ for the best-fit parameters: STv-HMF has 
$\chi^2_{ST\rm{v}}=1472\times10^{-3}$ and STc-HMF $\chi^2_{STc}=4325\times10^{-3}$. Therefore we can conclude that lower $p$ values can improve the agreement between the ST-HMF template and E-HMF at low masses, both for $\delta_c(z)$ and $\delta_{\rm v}(z)$ thresholds. But at high masses 
$\delta_{\rm v}(z)$ is the main factor providing a better agreement between ST template and E-HMF.

It is important to note that large disagreements between any two distinct HMFs models is inevitable at the high-mass tail because any small difference between the parameters appearing in the exponential part will greatly amplify the differences at these mass scales. However, the actual consequences of this fact are usually not so drastic because high-mass objects are rare and demand very large simulations to be properly sampled, resulting in HMFs with decreasing accuracies with mass. Therefore, possible cosmological constraints and comparisons between HMFs have limited significance when considering the most massive halos at each epoch. The main message in the analysis portraited in figure \ref{fig:ST-error} is the improved agreement between STv-HMF and the E-HMF with respect to the other versions of the ST template that use $\delta_c$. As we will further discuss in the next section, given this extended compatibility, the use of the STv-HMF is a good alternative to extrapolate the applicability of the E-HMF to models more complex than $\Lambda$CDM.

\section{Beyond $\Lambda$CDM models\label{sec:Beyond-CDM-models}}

As a further consistency check of the use of $\delta_{{\rm v}}$ as
the threshold for halo formation in the ST template, we will analyse
the level of agreement between E, ST and STv-HMFs when computing the
impact of smooth DE models with $w\neq-1$. HMFs for smooth DE with
varying equation of state parameter have been studied by \cite{OndaroMallea2021},
where the effects were modelled by an explicit dependence on the matter
growth rate, $f=d\ln \delta_m/d\ln a$. However, as shown in \cite{Euclid:2022dbc},
this HMF has an overall worse agreement with the E-HMF than the ST-HMF.
It is also discussed that the dependence on $f$ can be reparameterised
in terms of $\Omega_{m}(z)$, which appears in equation (\ref{eq:a-parametrization}),
making use of growth index parametrization $f=\Omega_{m}^{\gamma}(z)$.
Therefore, although the E-HMF has been fitted for the $\Lambda$CDM
model, one can expect that the impact of $w\neq-1$ can be consistently
captured by the explicit $\Omega_m(z)$ dependence\footnote{In a private communication, Tiago Castro mentioned that an upcoming paper will show that the E-HMF can describe, within a few percent accuracy, the halo abundances in DE models with equation of state described by the $w0wa$ parametrization.}, and it will be used here as a basis of comparison. 

A common technique used to extrapolate the validity of a fiducial
HMF, numerically calibrated with simulations for a specific cosmological
model, is to compute the comoving number density of halos as
\begin{equation}
\frac{dn}{dM}|_{\text{mod}}=R_{n}\frac{dn}{dM}|_{\text{fid}}\,,\label{eq:HMF-correction}
\end{equation}
where
\begin{equation}
R_{n}=\frac{dn/dM|_{\text{mod}}}{dn/dM|_{\text{ref. mod}}}\,,\label{eq:ratio-dn-dM}
\end{equation}
is computed with an HMF for which some analytical model can be used
to describe the features of the cosmological model under consideration
(indicated by the `mod' subscript) and its counterpart for the reference
model (`ref. mod' subscript, $\Lambda$CDM in our case), see, e.g.,
\cite{LoVerde2008,Creminelli2010,Batista:2013oca,Batista2017}.

In order to explore the impact of smooth DE models beyond $\Lambda$,
we compute how a particular HMF for $w\neq-1$ changes with respect
to the one for the $\Lambda$CDM model. As explained earlier, the
E-HMF should provide a good representation for them. In this case,
one would not need to compute $R_{n}$, and the E-HMF could be used
directly with the quantities associated with the new model. In what
follows, we also compute $R_{n}$ for the E-HMF as a reference for
comparison with the results for ST and STv-HMFs. In the following
examples, we assume the $w_{0}w_{a}$ parametrization for the equation
of state parameter (EoS) \cite{Chevallier:2000qy,Linder:2002et}
\begin{equation}
w\left(a\right)=w_{0}+w_{a}\left(1-a\right)\,.
\end{equation}

\subsection{Constant EoS}

\begin{figure}
\centering{}\includegraphics[scale=0.5]{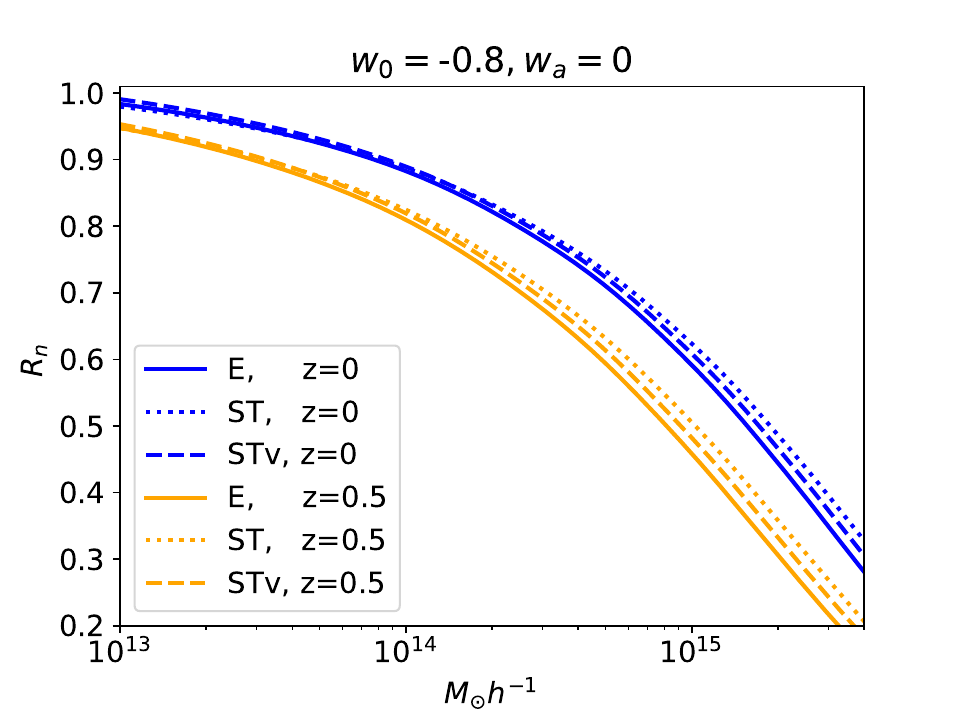}\includegraphics[scale=0.5]{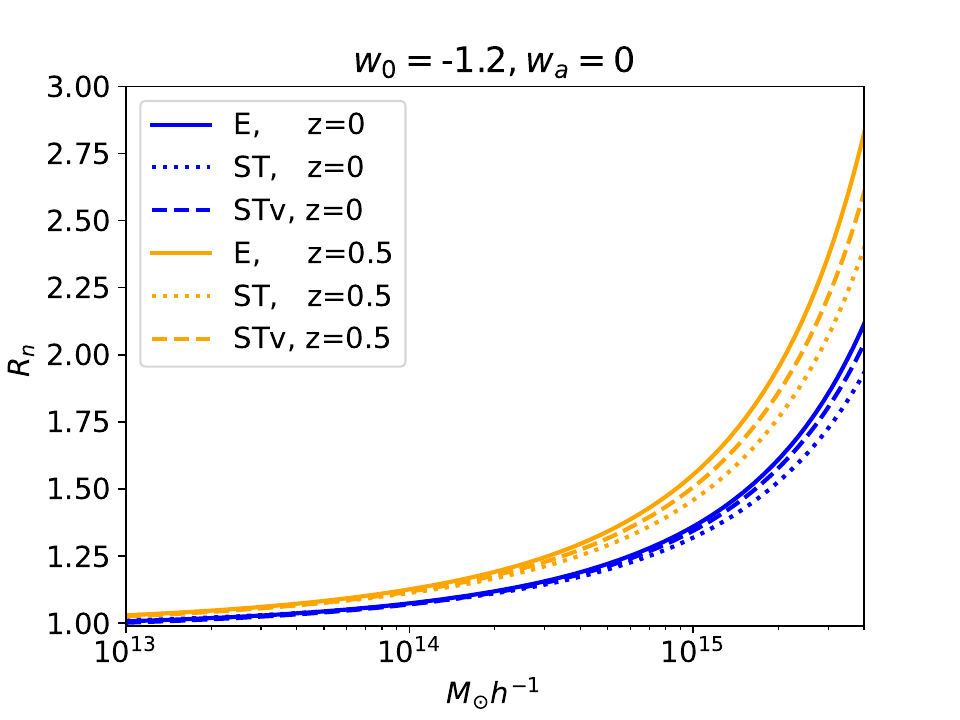}
\caption{$R_{n}$, defined in (\ref{eq:ratio-dn-dM}), for STv, ST and E-HMFs
(indicated in the plot legends) for dark energy models with $w_{0}=-0.8$
(left panel) and $w_{0}=-1.2$ (right panel) at $z=0$ (blue lines)
and $z=0.5$ (orange lines). In the left panel, the disagreement between $R_n$ from STv and E-HMF is below $5\%$ for    
$M<2\times10^{15} M_{\odot}h^{-1}$ at $z=0$ and $M<9.5\times10^{14} M_{\odot}h^{-1}$ at $z=0.5$. In the right panel, the disagreement between $R_n$ from STv and E-HMF is below $5\%$ for $M<2\times10^{15} M_{\odot}h^{-1}$ at $z=0$ and $M<9.5\times10^{14} M_{\odot}h^{-1}$ at $z=0.5$.}
\label{fig:HMF-diff-smooth} 
\end{figure}

In figure \ref{fig:HMF-diff-smooth} we show $R_{n}$ for ST, STv
and E-HMF for $w_{0}=-0.8$ and $w_{0}=-1.2$, with $w_{a}=0$ in
both cases. As can be seen, all the HMFs agree qualitatively, showing
a decrease of the comoving number density of halos for $w_{0}=-0.8$
and a increase for $w_{0}=-1.2$. We also note that STv-HMF has an
overall better agreement with E-HMF than ST-HMF, especially for high
masses. Therefore, we can conclude that the use of $\delta_{{\rm v}}$
in ST template also provides better agreement with E-HMF for smooth
DE models with $w_{0}\neq-1$ and $w_{a}=0$. Moreover, it is also
evident that the impact of $w_{0}$ in ST-HMF is the smallest at high
masses. This happens because ST-HMF effective threshold is essentially
constant, so the impact of distinct DE models is captured only by
the linear evolution of perturbations, encoded in $\sigma(M,z)$.

An important feature to note in these examples is the clear impact
of phantom ($w<-1$) and non-phantom EoS ($w>-1$) with respect to
$\Lambda$CDM. For $w_{0}=-0.8$, the DE fluid has more energy density
than $\Lambda$ at high-$z$, consequently $\Omega_{m}\left(z\right)$
is lower and the growth of matter fluctuations is relatively suppressed.
As a consequence, we always have $R_{n}<1$. The opposite occurs for
the phantom case of $w_{0}=-1.2$.

\subsection{Time varying EoS}

Although the impact of constant EoS is very clear, and all the three
HMFs agree qualitatively about it, the situation for time-varying
EoS can be more complex. As an example, we will analyse two pairs
of $\left(w_{0},w_{a}\right)$ values, whose growth history changes
in time with respect to the $\Lambda$CDM one. For $p_{1}=\left(w_{0}=-0.9,w_{a}=-0.3\right)$
the EoS is phantom at high-$z$ and non-phantom at low-$z$. For $p_{2}=\left(w_{0}=-1.104,w_{a}=0.3\right)$,
the converse holds. It is interesting to note that the first kind
of EoS evolution is supported by the recent DESI-BAO analysis \cite{Adame2024}. 

\begin{figure}
\centering{}
\includegraphics[scale=0.48]{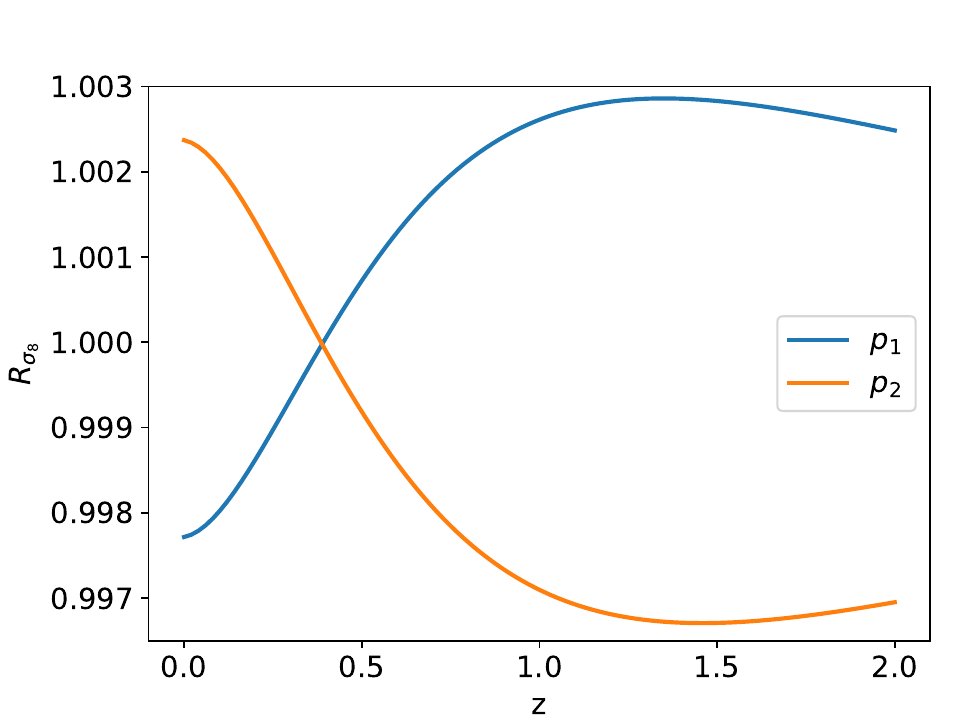}\includegraphics[scale=0.48]{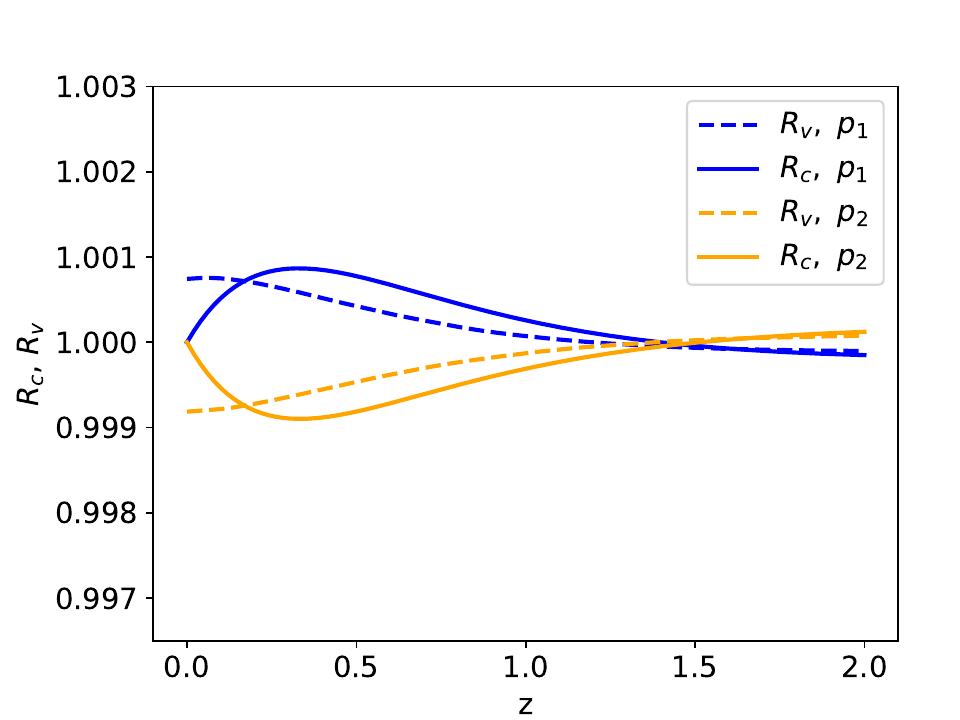}
\caption{Left panel: evolution of $R_{\sigma_{8}}$, defined in (\ref{eq:ratio-sigma}),
for $p_{1}=\left(w_{0}=-0.9,w_{a}=-0.3\right)$ and $p_{2}=\left(w_{0}=-1.104,w_{a}=0.3\right)$. Right panel: evolution of $R_{c}$ and $R_{{\rm v}}$, defined in (\ref{eq:Rc-Rv}), for the same pairs of parameters.}
\label{fig:ratio-sigma}
\end{figure}

Let us first analyse the relative linear growth history with respect
to $\Lambda$CDM. In the left panel of figure \ref{fig:ratio-sigma}, we show the ratio
between the evolutions of $\sigma_{8}\left(z\right)$, given by
\begin{equation}
R_{\sigma_{8}}=\frac{\sigma_{8}\left(z\right)_{{\rm \text{mod}}}}{\sigma_{8}\left(z\right)_{\Lambda\text{CDM}}}\,.\label{eq:ratio-sigma}
\end{equation}
For $p_{1}$, $\sigma_{8}(z)$ is higher than in $\Lambda$CDM at high-$z$.
At lower $z$, the EoS becomes non-phantom and there is less growth
with respect to $\Lambda$CDM. The opposite happens for $p_{2}$.
In both cases, this relative transition in the growth behaviour occurs
at $z_{t}\simeq0.38$ (the value of $w_{0}$ in $p_{2}$ was adjusted
to get approximately the same $z_{t}$ as for $p_{1}$).

Prior to $z_{t}\simeq0.38$, the growth histories of the two examples
are clearly enhanced or diminished with respect to $\Lambda$CDM.
It is not expected that the nonlinear evolution encoded in the effective
threshold of HMFs should present a distinct relative behaviour as
seen for $R_{\sigma_{8}}$. Therefore, at $z>z_{t}$, the ratio between
halo number densities, equation (\ref{eq:ratio-dn-dM}), should preserve
the relations seen in the left panel of figure \ref{fig:ratio-sigma}.

\begin{figure}
\centering{}\includegraphics[scale=0.5]{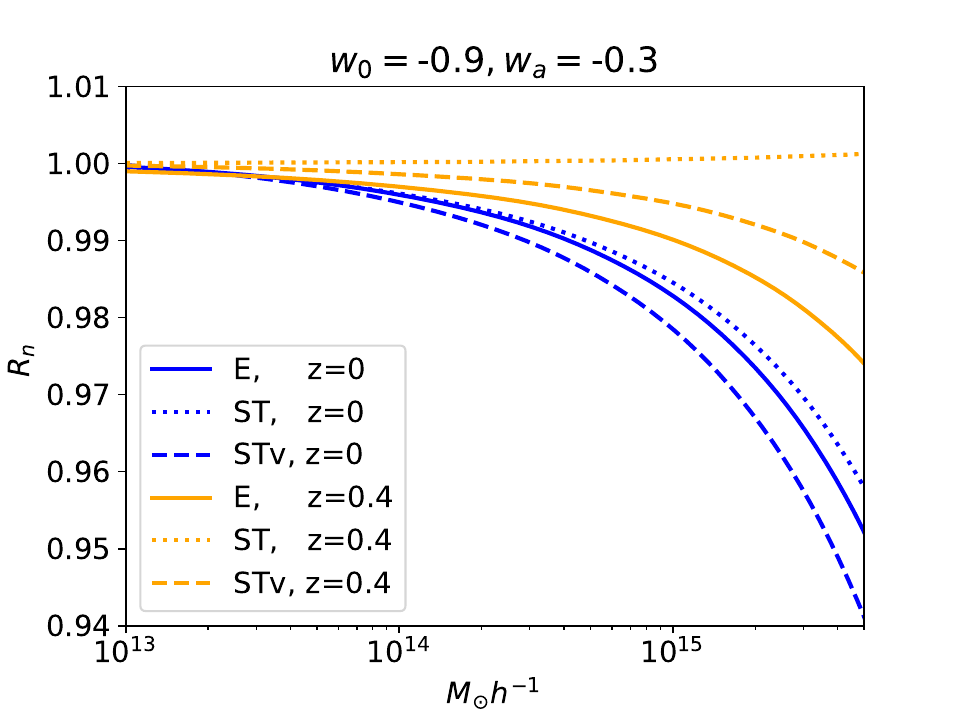}\includegraphics[scale=0.5]{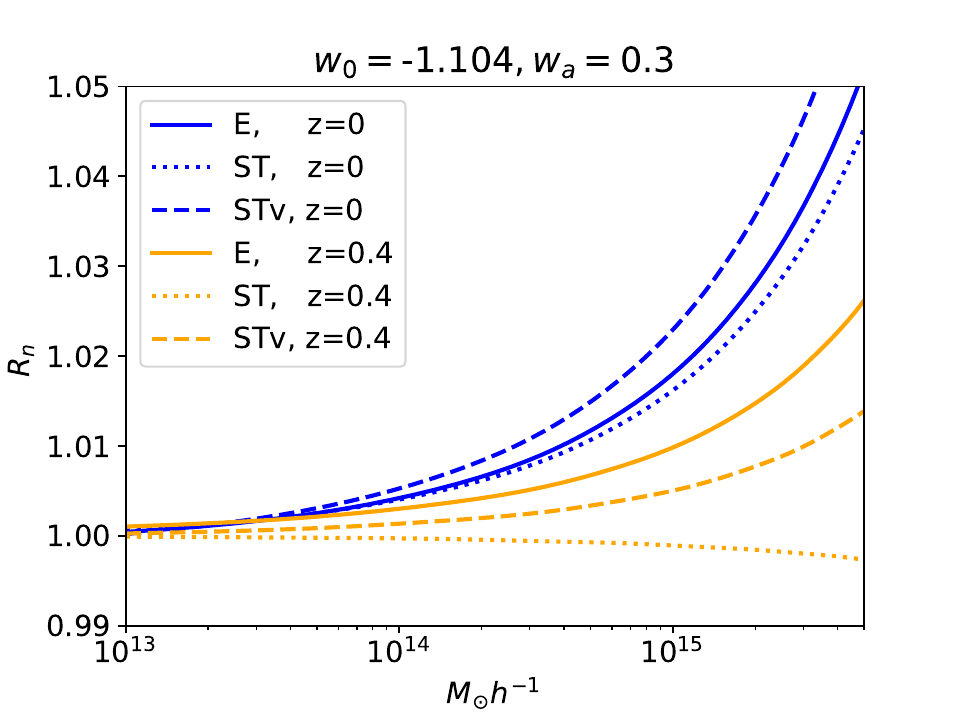}\caption{Ratios of STv, ST and E-HMFs (indicated in the plot legends) for DE
models with EoS parameters indicated at the plot titles at $z=0$
(blue lines) and $z=1$ (orange lines). In both panels, the disagreement between $R_n$ from STv and E-HMF barely exceeds $1\%$ for entire mass range shown, both  at $z=0$ and $z=0.4$.
\label{fig:HMF-diff-time-var-w}}
\end{figure}

However, these relations between models based on the linear growth
history is not always satisfied by HMFs. In figure \ref{fig:HMF-diff-time-var-w}
we show $R_{n}$ for the two models considered in this subsection.
As can be seen, at $z=0.4$, slightly before $z_{t}\simeq0.38$, only
ST-HMF presents the expected behaviour: $R_{n}>1$ for $p_{1}$ and
$R_{n}<1$ for $p_{2}$. This is easy to understand, given that the
effective threshold in the ST-HMF depends on $\delta_{c}(z)$, which
is nearly constant, the main impact in $R_{n}$ is due to $\sigma\left(M,z\right)$.
For STv and E-HMF, the effective thresholds vary in time, but in a
manner not totally consistent with the variation of $\sigma_{8}\left(z\right)$. 

As can be seen in figure \ref{fig:HMF-diff-time-var-w}, at $z=0$, all three
HMFs agree qualitatively. We have also verified that, for both $p_{1}$
and $p_{2}$, STv-HMF has the expected $R_{n}$ for $z>0.46$ and
E-HMF for $z>0.51$. For $z>0.51$, all three HMFs agree about the
impact of EoS in number density with respect to $\Lambda$CDM. We
also have checked the behaviour of $R_{n}$ given by the HMF proposed
in \cite{OndaroMallea2021}, finding very similar results to the ones
observed for ST-HMF. This is expected because the HMF proposed in
\cite{OndaroMallea2021} considers multiplicative corrections on the
ST template.

In order to clearly see that this undesirable effect is caused by
the effective threshold, in the right panel of figure \ref{fig:ratio-sigma}, we show the redshift
evolution of the ratios
\begin{equation}
R_{c}=\frac{D_{c\,\text{mod}}(z)}{D_{c\,\Lambda}(z)}\,\,\,\text{and }R_{{\rm v}}=\frac{\delta_{{\rm v\,}\text{mod}}(z)}{\delta_{{\rm v}\,\Lambda}(z)}\,,
\label{eq:Rc-Rv}
\end{equation}
where $D_{c}$ is the time-dependent part of the effective threshold
in the E-HMF, defined in (\ref{eq:definition-Dc-Dv}). At high-$z$,
both quantities are lower than unity for $p_{1}$ and higher than
unity for $p_{2}$, as expected for enhanced and diminished growth
histories with respect to $\Lambda$CDM, respectively. The inversion
in the relative behaviour occurs much earlier than the one related to
$\sigma_{8}\left(z\right)$, shown in the left panel of figure \ref{fig:ratio-sigma}.
After this inversion, $R_{\rm v}$ remains above or lower the unity, whereas
$R_{c}\rightarrow1$ as $z\rightarrow0$. Moreover, $R_{c}$ reaches
higher or lower values than $R_{\rm v}$, with a stronger time variation.

Although these differences with respect to $\Lambda$CDM are small,
when compared to $R_{\sigma_{8}}$ variations, the behaviour of $R_{c}$
and $R_{{\rm v}}$ do not totally agree with $R_{\sigma_{8}}$ shown
in figure \ref{fig:ratio-sigma}. Inversions in $R_{c}$ and $R_{{\rm v}}$
occur much earlier than the ones for $R_{\sigma_{8}}$. This would
indicate that the relative inversion in the effective threshold, which
encodes the nonlinear evolution, can anticipate the one associated
with the linear evolution, encoded in $\sigma_{8}\left(z\right)$,
which looks highly unlikely to be true. 

Moreover, $R_{c}$ shows a second problem. Given that the redshift
evolution of the effective threshold in the E-HMF mainly depends on
$\Omega_{m}^{a_{z}}(z)$, for fixed $\Omega_{m0}$, $R_{c}$ for different
models have the same value at $z=0$. This looks unnatural and is
in disagreement with behaviour $R_{\sigma_{8}}$ shown in figure \ref{fig:ratio-sigma}.
Such behaviour would indicate that, although a specific model has
more/less linear growth than $\Lambda$CDM at low-$z$, the effective
threshold does not respond to this increase/decrease of growth at
low-$z$. This issue might indicate that the parameter $a_{z}$ in
the E-HMF should have some dependence on the EoS. On the other hand,
$R_{{\rm v}}$ reproduces the behaviour of $R_{\sigma_{8}}$ at low-$z$,
what looks more natural.

However, it is important to note that these theoretical issues with
the models with time-varying EoS might not be problematic when confronted
with observational or simulated data. For the two cases analysed,
the undesirable behaviour of $R_{n}$ occurs at intermediate redshifts
($0.4\lesssim z\lesssim0.5$) with more impact for the most massive
halos, which are not so abundant early on. In general, and as expected,
the differences in $R_{n}$ between HMFs increase at high masses,
but exceed $1\%$ only for rare objects, e.g., $M>2\times10^{15}M_{\odot}h^{-1}$
at $z=0$ or $M>1\times10^{15}M_{\odot}h^{-1}$ at $z=0.4$. Moreover,
the actual number of observed clusters also depend on the comoving
volume, which is analytically determined and can have large variations
with the EoS.

\section{Conclusions \label{sec:Conclusions}}

We have analysed the redshift dependence of the effective threshold
in the E-HMF, proposed in \cite{Euclid:2022dbc}, showing it is qualitatively
equivalent to the virialization threshold, $\delta_{{\rm v}}(z)$, and
incompatible with the evolution of the collapse threshold, $\delta_{c}(z)$,
which has been used in many HMFs so far. This finding indicates that
the abundance of halos in the Universe is better described by the
physics of virialization, encoded in $\delta_{{\rm v}}(z)$, than physics
of collapse, associated with $\delta_{c}(z)$. 

We found that the redshift-dependent part of effective threshold in the E-HMF, $\Omega_{m}^{a_{z}}(z)\delta_{c}^{2}(z)$,
can be accurately represented by a simple power-law function of the
virialization threshold, given by $\alpha\delta_{{\rm v}}^{\beta}(z)$,
which is also robust against variations of $\Omega_{m0}$, showing
that the natural dependence of the threshold on $\delta_{{\rm v}}(z)$
is not a coincidence for a particular expansion history in the $\Lambda$CDM
model. We found $\beta\simeq4.37$, which shows that the effective
threshold in the E-HMF grows faster than $\delta_{{\rm v}}^{2}(z)$ at
low-$z$. We interpret this as an indication that a more realistic
model for the nonlinear dynamics and virialization should provide
faster growing $\delta_{{\rm v}}(z)$, in such a way that effective threshold
is proportional to the square of this new function, as demanded by
the analytical formulation of the HMF.

Exploring these findings, we have shown that a refitted ST-HMF with
the substitution $\delta_{c}(z)\rightarrow\delta_{{\rm v}}(z)$ has better
agreement with the E-HMF, which is also robust when varying $\Omega_{m0}$.
Interestingly, this refit produce parameters $a$ and $p$ closer to $1$ and $0$, respectively, making the STv-HMF more similar to the PS-HMF. Therefore, the STv-HMF is less dependent on ad-doc parameters, while providing better concordance with the E-HMF then the ST template with $\delta_{c}(z)$.  In this sense, the STv-HMF preserves the universality features of the 
ST template with improved accuracy.

We also analysed the case of smooth DE with constant and time-varying
EoS. For constant $w$, all the three HMF under consideration (ST,
STv and E-HMF) agree qualitatively about the relative abundance number
density of halos with respect to $\Lambda$CDM. In this case, the
changes in $\sigma\left(M,z\right)$ are more relevant than the ones
in the effective threshold. In this scenario, the STv-HMF has a better
agreement with E-HMF than ST-HMF for both phantom and non-phantom
EoS in the mass range under consideration. 

The case of time-varying EoS is more complex. Considering two examples
of the $w_{0}w_{a}$ parametrization that transit from phantom to
non-phantom regimes, and vice-versa, we showed that the impact of
time-varying EoS in the STv and E-HMFs is not totally consistent with
the linear growth history encoded in $\sigma_{8}\left(z\right)$.
Moreover, in these examples, the time-dependent part of the effective
threshold in E-HMF approaches the same value as in $\Lambda$CDM at
$z=0$, in qualitatively disagreement with the behaviour of $\sigma_{8}\left(z\right)$
for distinct models. However, given the small deviations from the
expected behaviour and that their impact occurs mostly at the high-mass
tail of the HMFs, it is likely that the observational consequences
of this issue are very small, or even negligible.

Our main conclusion is that the virialization threshold is a sensible
basic model for the description of the time-dependence of the effective
threshold for smooth DE models. Given that $\delta_{{\rm v}}$ is
computed by a physical model, in principle, it should provide a reliable
manner to determine the impact models beyond smooth DE in halo abundances,
such as clustering DE, or even for the scenarios of 
modified gravity and models beyond cold dark
matter. Furthermore, at least in the case of clustering DE, it is
not numerically possible to compute $\delta_{c}$ \cite{Batista:2022ixz},
thus the use of $\delta_{{\rm v}}$ is essential to explore the behaviour
of such models with semi-analytic HMFs.

\acknowledgments{
RCB thanks Valerio Marra for useful discussions and Tiago Castro for
critical analysis of several ideas in this work.} 

\appendix
\section{ $\delta_{{\rm v}}$ fit for $\Lambda$CDM and a general code for its computation}

It might be helpful to determinate a representation of $\delta_{{\rm v}}$
in terms of $\delta_{{\rm c}}$ for the $\Lambda$CDM model, because
the latter has a simple and well-known fitting function, given by
\cite{Kitayama1996}. We proceed with the following representation
\begin{equation}
\delta_{{\rm v}}(z)=\alpha\delta_{{\rm c}}^{\beta}(z)
\label{eq:delta_vir_par_delta_c}
\end{equation}
and determine the constant values $\alpha$ and $\beta$. We assume
that $\delta_{{\rm v}}$ and $\alpha\delta_{c}^{\beta}$ match
in the EdS limit, so we have the analogous of equation (\ref{eq:alpha1-definition})
for $\alpha$ and
\begin{equation}
\beta=\frac{\ln\left(\delta_{\rm v}(z)/\delta_{{\rm v}\text{EdS}}\right)}{\ln\left(\delta_{c}(z)/\delta_{c\text{EdS}}\right)}\,.
\end{equation}
For $\Omega_{m0}=0.3$ and $0<z<2$, we have $-2.27>\beta>-2.55$.
We compute an effective value in the same way as in equation (\ref{eq:xi2}),
getting 
\begin{equation}
\alpha=5.4078\,\,\,\text{and }\beta=-2.3509\,.
\end{equation}

\begin{figure}
\centering{}\includegraphics[scale=0.6]{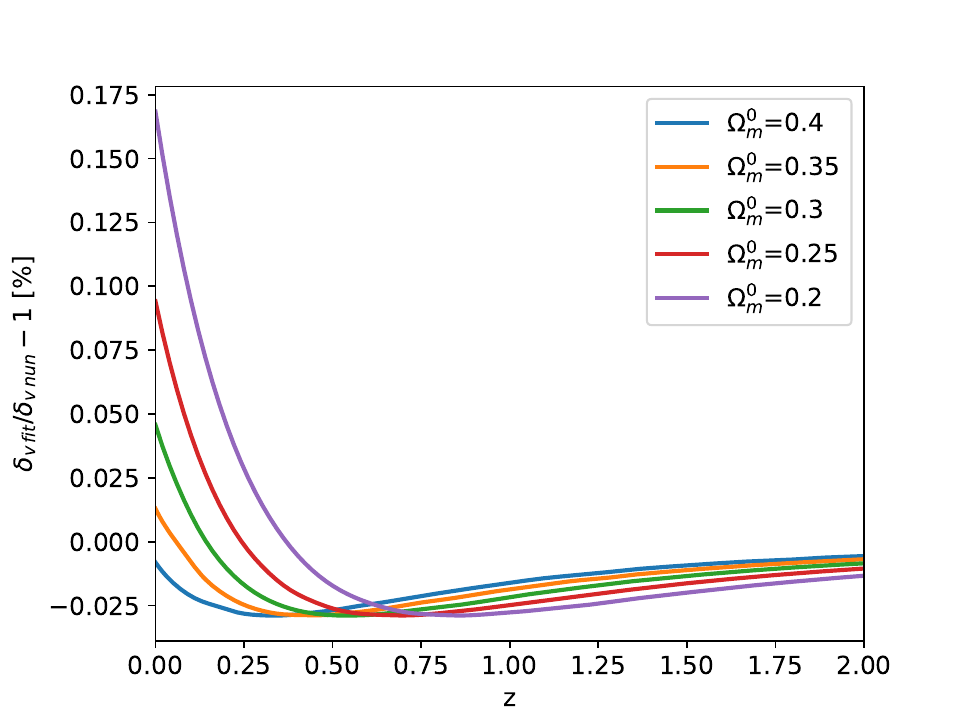}\caption{Percent difference between the representation of $\delta_{{\rm v}}$ given by (\ref{eq:delta_vir_par_delta_c}) and its numerical solution in the $\Lambda$CDM model for several values of $\Omega_{m0}$
indicated in the plot legend.\label{fig:nu_par_diff}}
\end{figure}

In figure \ref{fig:nu_par_diff}, we show the percent difference between
the numerical solution and the fit given by (\ref{eq:delta_vir_par_delta_c}). As can be seen,
this representation of $\delta_{{\rm v}}$ is very precise and robust
when varying $\Omega_{m0}$, with error below $0.17\%$ for $0.2\le\Omega_{m0}<0.4$.

Moreover, a code to compute the evolution of matter fluctuations
in the SC model and the determination of the virialization
threshold, $\delta_{{\rm v}}$, for smooth DE models described by the $w_{0}w_{a}$
parametrization is publicly available at \href{https://github.com/roncab/scollas}{https://github.com/roncab/scollas}.

\bibliographystyle{/home/rbatista/work/references/JHEP}
\bibliography{/home/rbatista/work/references/referencias}

\end{document}